\documentclass[notoc]{JHEP3}
\usepackage{amsmath,amssymb}
\usepackage{graphicx}

\newcommand{\be}{\begin{equation}}
\newcommand{\ee}{\end{equation}}
\newcommand{\bea}{\begin{eqnarray}}
\newcommand{\eea}{\end{eqnarray}}
\newcommand{\beann}{\begin{eqnarray*}}
\newcommand{\eeann}{\end{eqnarray*}}

\newcommand{\ba}{\begin{array}}
\newcommand{\ea}{\end{array}}

\newcommand{\cL}{{\cal L}}

\title{ R**2 correction to BMPV black hole entropy from Kerr/CFT correspondence 
}

\author{Hirotaka Hayashi$^{1}$ and Ta-Sheng Tai$^2$\\

$^1$School of Physics\\ Korea Institute for Advanced Study\\ 
Seoul 130-722, Korea\\

$^2$Interdisciplinary Graduate School of Science and Engineering\\
Kinki University\\ 
Osaka 577-8502, Japan\\

\\
\\
{\tt hayashi} {\rm at} {\tt kias.re.kr},
{\tt tasheng} {\rm at} {\tt alice.math.kindai.ac.jp}
}

\preprint{ KIAS-P11083
           }

\abstract{
Following Kerr/CFT correspondence, we compute the entropy of five-dimensional supersymmetric rotating 
BMPV black holes. 
We successfully reproduce Iyer-Wald formula in the presence of Gauss-Bonnet terms 
from the viewpoint of microscopic CFT. 
This further supports the higher-derivative version of 
Kerr/CFT prescription proposed in arXiv:0903.4176 for 
four-dimensional extremal Kerr black holes.}

\begin{document}


\section{Introduction}
\label{sec:intro}

A microscopic entropy counting programme called Kerr/CFT correspondence was proposed by Guica et al. in \cite{0809.4266}. They have examined 4D extremal Kerr black holes whose near-horizon geometry has $SL(2,R) \times U(1)$ isometry. 
Further extension can be found in  \cite{0810.2620}-\cite{Nakamura:2011gn}. 
Remarkably, the tree-level black hole entropy $S_0$ predicted by 
Bekenstein-Hawking area law was 
reproduced by applying 
Cardy's formula to the dual 2D CFT: 
\begin{eqnarray} 
\label{ca}
S_0=\frac{\pi^2 }{3}c T_{FT}
\end{eqnarray} 
where $T_{FT}$ denotes Frolov-Thorne temperature \cite{FTT}. 

Let us momentarily clarify the nature of Kerr/CFT correspondence. 
Compared with the BTZ/CFT case \cite{BH} pioneered earlier by Brown and Henneaux,%
\footnote{See Saida and Soda \cite{Saida:1999ec} for the extension of Brown-Henneaux to the inclusion of 
Gauss-Bonnet terms.} here one takes into account the asymptotic symmetry group (ASG) of 
the near-horizon black hole geometry instead of its asymptotically-far one. 
We will adopt the terminology $asymptotic$ Killing vector (instead of Killing vector) 
especially for the former case under consideration. 
In addition, the central charge $c$ in \cite{BH} did arise from the enhanced $SL(2,R)_L \times SL(2,R)_R$, 
the isometry of the asymptotic 3D BTZ (or $AdS_3$), to two copies of (chiral and anti-chiral) Virasoro algebras. 
In contrast to this, $c$ in \eqref{cc} has a rather different nature. That is, given the above asymptotic Killing vector field like \eqref{killing} 
infinitely many Fourier modes $\zeta_n$ are identified with generators $L_n$ 
of Virasoro algebra whose central charge \eqref{cc} is 
determined completely by the near-horizon metric and \eqref{killing}. 

In this Letter, within the framework of Kerr/CFT correspondence 
we focus on the entropy of 5D BMPV black holes 
in the presence of $R^2$-curvature corrections.%
\footnote{See also 
\cite{Castro:2008ys, Banerjee:2008ag, Dabholkar:2010rm} for the 
derivation of subleading corrections from other viewpoints.} 
In \cite{Krishnan:2009tj, 0903.4176}, this direction has been explored in the context of 
4D extremal Kerr ones. We still rely on Cardy's formula \eqref{ca} but modify the central charge $c$ due to the inclusion of Gauss-Bonnet terms. 
It is in \cite{0111246, 0708.2378, 0708.3153} that the explicit expression of $c$ in the presence of higher-derivative corrections has been spelt out. 
Combined with Frolov-Thorne temperature analyzed  
in \cite{Azeyanagi:2008dk, 0812.4440} for BMPV black holes, we found that \eqref{ca} coincides perfectly 
with Iyer-Wald formula \cite{1,2,3}.

We organize this Letter as follows. In Section \ref{sec:BMPV}, we describe 
some basic aspects of BMPV black holes. 
Iyer-Wald entropy formula 
in the presence of Gauss-Bonnet terms is also reviewed. In Section \ref{sec:kerr}, we evaluate BMPV entropy 
microscopically via Kerr/CFT correspondence as stated above. 
We end up this Letter with a summary in Section \ref{sec:sum}. 

\section{ The BMPV black hole entropy}
\label{sec:BMPV}

The BMPV black hole was first constructed 
in \cite{Breckenridge:1996is}. 
As shown by Kallosh et al. \cite{Kallosh:1996vy}, it can get embedded in 5D $\mathcal{N}=2$ 
supergravity coupled to one vector multiplet, and 
preserves one-half supersymmetry.
The metric of the BMPV black hole is%
\begin{eqnarray}
ds^{2} = -\left(1-\frac{\mu}{r^{2}}\right)^{2}dt^{2} + \frac{dr^{2}}{\left(1-\frac{\mu}{r^{2}}\right)^{2}} - \frac{\mu a}{r^{2}}\left(1-\frac{\mu}{r^{2}}\right)\sigma_{3}dt -\frac{\mu^{2}a^{2}}{4r^{4}}\sigma_{3}^2 + \frac{r^{2}}{4}d\Omega_{3}^{2}
\label{eq:BMPV}
\end{eqnarray}
where 
\begin{eqnarray}
&&\sigma_{3} = d\varphi + \cos\theta d\psi,\qquad d\Omega_{3}^{2} = d\theta^{2} +\sin^{2}\theta d\psi^{2} + \sigma_{3}^{2},\nonumber\\
&&0 \leq \theta < \pi, \qquad 0 \leq \psi < 2 \pi, \qquad 0 \leq \varphi < 4 \pi \nonumber.
\end{eqnarray}
Here, $\frac{1}{4}d\Omega_{3}^{2}$ is the line element of a unit $S^{3}$. 

The lowest scalar component of the vector multiplet can be set 
to some constant. Because of this fact, 
the gaugino equation 
implies a vanishing gauge field in the vector multiplet. On the other hand, the graviphoton gauge potential in the graviton multiplet is 
\begin{eqnarray}
A =B(r)dt + C(r)\sigma_{3}, \label{gauge}
\end{eqnarray}
where 
\begin{eqnarray}
B(r) = \frac{\sqrt{3} \mu}{2r^{2}}, \qquad C(r) = - \frac{\sqrt{3} \mu a}{4 r^{2}}. \nonumber
\end{eqnarray}

The conserved charges of the BMPV black hole \eqref{eq:BMPV} can be measured at the asymptotic spatial infinity 
according to \cite{Myers:1986un}. Thus, 
the conserved angular momentum along $\varphi$ and the electric charge measured at asymptotic infinity of the BMPV black hole \eqref{eq:BMPV} are 
\begin{eqnarray}
J \equiv  J_{\varphi} &=& -\frac{1}{16 \pi G_5}\int_{\infty} \ast \nabla \xi^{\varphi} = -\frac{\pi a \mu}{4 G_5},\label{angular}\\
Q &=& \frac{1}{4 \pi G_5} \int_{\infty} \ast dA = -\frac{\sqrt{3} \pi\mu}{2 G_5}
\end{eqnarray}
where $\nabla_{\mu}(\xi^{\varphi})_{\nu}dx^{\mu} \wedge dx^{\nu}$ is abbreviated to 
$\nabla \xi^{\varphi}$ 
with $\xi^{\varphi}$ being a Killing vector field $\frac{\partial}{\partial\varphi}$. Here, 
$x^{\mu}$ represents the coordinates $(t, r, \theta, \psi, \varphi)$. We have also expressed the 5D Newton constant as $G_5$. The ADM mass of 
the BMPV black hole, 
$M_{ADM} = \frac{3\pi\mu}{4G_5}$, 
is proportional to $Q$ as required by supersymmetry. One can also see that there is no angular momentum along $\psi$.


Another realization of the BMPV black hole is through 
compactifying the ten-dimensional Type IIA string theory on a Calabi-Yau threefold $X$ \cite{Gaiotto:2005gf}. 
Wrapped on $X$ is a D0-D2-D4-D6 brane system with 
$( q_0, q_A, p^A, p^0)$ with $A = 1, \cdots, h^{1,1}(X)$ indicating their RR charges (or the numbers).
In the case of the BMPV black holes, one has $( q_0, q_A, p^A, p^0)=( q_0, q_A, 0, 1)$. 
Due to one single D6-brane, the eleven-dimensional M-theory lift of this brane system 
amounts to placing the resultant 5D charged and rotational black hole 
at the center of a Taub-NUT space $TN_4$ (in order to preserve supersymmetry). Notify that $q_0 \propto J_{\varphi}$ is the spin over the $S^1$ bundle of $TN_4$. This stringy construction 
makes natural both the appearance of $S^3 \subset TN_4$ in 
\eqref{eq:BMPV} and the statement of no angular momentum along $\psi$. 
In addition, $q_A$ is related to $Q$ by $q_A = 3 Q D_{ABC} Y^B Y^C$ where 
 $D_{ABC}$ is the triple intersection number of $X$ and $Y^A$ stands for horizon values of scalar components in ${\cal N}=2$ vector multiplets. 
They are normalized by $D_{ABC}Y^A Y^B Y^C =1$. 

In order to consider the near-horizon limit, let $\tilde{r} = r - \sqrt{\mu}$. 
In the near-horizon limit $\tilde{r} \rightarrow 0$, the BMPV black hole metric \eqref{eq:BMPV} becomes 
\begin{equation}
ds^{2} = -\left(\frac{2 \tilde{r}}{\sqrt{\mu}}\right)^{2} dt^{2} + \left(\frac{\sqrt{\mu}}{2 \tilde{r}}\right)^{2} d\tilde{r}^{2} - \frac{2 a}{\sqrt{\mu}}\tilde{r}\sigma_{3}dt + \frac{\mu -a^{2}}{4}\sigma_{3}^{2} + \frac{\mu}{4}(d\theta^{2} + \sin^{2}\theta d\psi^{2}). 
\label{eq:BMPV_horizon}
\end{equation}
Obviously, when $J=0$ ($a=0$) \eqref{eq:BMPV_horizon} reduces to simply 
a direct product:  
$AdS_2 \times S^3$. 
For generic $J\ne 0$ cases, the bosonic isometry group of 
\eqref{eq:BMPV_horizon} gets broken down to $SU(1,1) \times SU(2) \times U(1)$ \cite{Gauntlett:1998fz}. By completing the square term $-(\frac{2\tilde{r}}{\sqrt{\mu}}dt+ \frac{a}{2}\sigma_3)^2$, 
the near-horizon topology of the BMPV black hole becomes 
$AdS_2$ fibered over $S^3$ \cite{Li:2006uq}. This may be a 
sign of the existence of certain putative dual 2D chiral CFT. 
From \eqref{eq:BMPV_horizon} it is easily seen that 
the tree-level entropy is 
\begin{equation}
S_{0} = \frac{{\rm Area}}{4 G_5} = \frac{\pi^2}{2 G_5} \mu \sqrt{\mu - a^{2}}. \label{entropy}
\end{equation}


\subsection{Entropy from the Iyer-Wald formula}
Let us see how $R^2$-curvature corrections modify \eqref{entropy}. 
When it comes to the Gauss-Bonnet terms, the result 
has actually been calculated in \cite{Guica:2005ig}. 
Our goal here is to reproduce the known result and meantime set up some notational conventions to be used 
in the next section.

The 5D ${\mathcal N}=2$ supergravity can arise from 
compactifying 11D M-theory on Calabi-Yau threefolds $X$ where the number of vector multiplets is $h^{1,1}(X)$. 
The resultant Einstein-frame low-energy effective action may necessarily contain higher-derivative corrections 
\cite{hep-th/9602102, hep-th/9707013} such as the familiar Gauss-Bonnet term: 
\begin{align}
\begin{split}
\delta {\mathcal L}= 
 \xi (R_{\mu \nu \rho \sigma} R^{\mu \nu \rho \sigma} -4 R_{\mu \nu}R^{ \mu \nu} + R^2)
\label{higher}
\end{split}
\end{align}
where $\xi$ is 
is proportional to $c_{2 A}Y^{A}$. 
As mentioned before, $Y^{A}$ denotes the 
scalar component in vector multiplets evaluated at the horizon. $c_{2 A}$ is given by  
\begin{align}
\begin{split}
c_{2 A}=\int_X c_2(X) \wedge \omega_{A} 
\end{split}
\end{align}
where $c_{2}(X)$ is the second Chern class of $X$ and $\omega_A \in H^{1,1}(X)$. What appeared in \eqref{eq:BMPV}-\eqref{gauge} 
correspond to the situation where one considers a special 
family of Calabi-Yau threefolds subject to 
$h^{1,1}(X) = 1$. We will concentrate only on this case. 

We call $S_1$ the correction to the tree-level entropy $S_0$ due to the higher-derivative corrections $\delta {\mathcal L}$. 
The correction $S_1$ to the tree-level entropy $S_0$ is computed by 
means of the Iyer-Wald formula\footnote{Certainly, 
the tree-level entropy formula \eqref{entropy} can also be recovered 
by considering the Einstein-Hilbert term for the Lagrangian density, namely 
$Z_{\mu \nu \rho \sigma} = {\partial(\sqrt{g} R)}/{\partial R^{\mu \nu \rho \sigma}}$.}
\cite{1,2,3}:  
\begin{align}
\begin{split}
\label{S1}
S_1= -2 \pi \int_{\Sigma} 
Z_{\mu \nu \rho \sigma}
\epsilon^{\mu \nu} \epsilon^{\rho \sigma}{\rm vol}(\Sigma),~~~~~~~~~
Z_{\mu \nu \rho \sigma}=\frac{\partial \delta {\cal L}}{\partial R^{\mu \nu \rho \sigma}} 
\end{split}
\end{align}
where ${\rm vol}(\Sigma)$ stands for the volume form over the horizon $\Sigma$. Here, $\epsilon^{\mu \nu}$ ($\epsilon^{\mu \nu}\epsilon_{\mu \nu}=-2$) denotes the binormal to $\Sigma$. 
Plugging \eqref{higher} into \eqref{S1}, we have 
\begin{equation}
Z_{\mu \nu \rho \sigma} = \xi\big(2R_{\mu \nu \rho \sigma} - 2(R_{\mu \rho} g_{\nu \sigma} - R_{\mu \sigma}g_{\nu \rho} + R_{\nu \sigma} g_{\mu \rho} - R_{\nu \rho}g_{\mu \sigma}) + R(g_{\mu \rho}g_{\nu \sigma}-g_{\mu \sigma}g_{\nu \rho})\big).
\label{Z}
\end{equation}
Note that the form of~\eqref{Z} arises according to the guideline that indices of $Z_{\mu \nu \rho \sigma}$ should obey 
\begin{equation}
Z_{\mu \nu \rho \sigma} = - Z_{\nu \mu \rho \sigma} = -Z_{\mu \nu \sigma \rho} = Z_{\nu \mu \sigma \rho} = Z_{\rho \sigma \mu \nu},
\end{equation}
as those of the Riemann tensor $R_{\mu \nu \rho \sigma}$ do.

In order to carry out the computation in \eqref{S1} 
with respect to the tree-level metric \eqref{eq:BMPV_horizon}, it may be easier to take an orthonormal basis instead of $dx^{\mu}$. An orthonormal basis $e^{\hat{a}} = e^{\hat{a}}{}_{\mu} dx^{\mu}, (\hat{a}=\hat{t}, \hat{r}, \hat{\theta}, \hat{\psi}, \hat{\varphi})$ of \eqref{eq:BMPV_horizon} is
\begin{eqnarray}
e^{\hat{t}} &=& \frac{2\tilde{r}}{\sqrt{\mu -a^2}} dt, \quad e^{\hat{r}} = \frac{\sqrt{\mu}}{2\tilde{r}} d\tilde{r}, \quad e^{\hat{\theta}} = \frac{\sqrt{\mu}}{2} d\theta,\nonumber \\
e^{\hat{\psi}} &=& \frac{\sqrt{\mu}}{2} \sin\theta d\psi, \quad 
 e^{\hat{\varphi}} = \frac{\sqrt{\mu -a^2}}{2}\left(d\varphi + \cos\theta d\psi - \frac{4\tilde{r}a}{\sqrt{\mu}(\mu-a^2)}dt\right). \label{frame2}
\end{eqnarray}
If one uses the orthonormal basis \eqref{frame2}, the non-zero components for 
the binormal $\epsilon^{\hat{a} \hat{b}}$ simply become $\epsilon^{\hat{t}\hat{r}} = -\epsilon^{\hat{r}\hat{t}} = 1$ and the others are zero. Then, the Iyer-Wald formula can be simply written by
\begin{equation}
S_{1} 
=-2 \pi \int_{\Sigma} (4Z_{\hat{t}\hat{r}\hat{t}\hat{r}})\; e^{\hat{\theta}} \wedge e^{\hat{\psi}} \wedge e^{\hat{\varphi}}.
\label{wald1} 
\end{equation}
When one explicitly inserts the near-horizon metric \eqref{eq:BMPV_horizon} to \eqref{wald1}, one obtains the correction $S_1$ to the tree-level entropy $S_0$ \eqref{entropy},
\begin{equation}
S_1 = 32 \pi^3 \xi \; \mu\sqrt{\mu -a^2} \frac{(3\mu + a^2)}{\mu^2}. 
\label{rere1}
\end{equation}

\section{Entropy from the Kerr/CFT correspondence}
\label{sec:kerr}

Having obtained \eqref{rere1} from the geometric viewpoint, we go to provide a statistical (or microscopic) derivation by means of 
the Kerr/CFT prescription. 
\subsection{Central charge of dual 2D CFT}

As shown in \cite{0809.4266}, because a suitable 
asymptotic boundary condition is imposed on the 
near-horizon metric there appears a series of 
vector fields which preserves them. We may call them \emph{asymptotic} Killing vector fields. 
One has to tell the difference between them and usual Killing vector fields as advocated in Section \ref{sec:intro}. 
The asymptotic boundary condition for the near-horizon metric \eqref{eq:BMPV_horizon} has been worked out in \cite{0812.4440}. 
It is preserved by the following vector field: 
\begin{align}
\begin{split}
\zeta_{n} = \sum_\alpha \zeta^\alpha_n \partial_\alpha =
- e^{-in \varphi} (\partial_\varphi + 
 in\tilde{r} \partial_{\tilde{r}}).
\label{killing}
\end{split}
\end{align}
$\zeta$'s satisfy the Witt algebra $i [\zeta_m, \zeta_n ]=(m-n)\zeta_{m+n}$. 

To yield the 
Virasoro algebra one has to add 
a central term.  The central extension can be accomplished by using an asymptotic charge associated to the asymptotic Killing vector field 
\eqref{killing}. 
The general formula for the asymptotic charge 
with respect to a given Lagrangian density including higher-derivative
terms has been obtained in \cite{0111246, 0708.2378, 0708.3153}: 
\begin{eqnarray}
c&=& 12i \Big\{ -2  \int_{\Sigma} 
\left[
\mathbf{ X}_{\alpha\beta} \mathcal{L}_{\zeta_n}
\nabla^\alpha \zeta^{\beta}_{-n}
+\left( \mathcal{L}_{\zeta_n} \mathbf{ X}\right)_{\alpha\beta}
\nabla^{[\alpha} \zeta^{\beta]}_{-n}
+ \mathcal{L}_{\zeta_n} \mathbf{W}_\alpha \zeta^{\alpha}_{-n}
\right]
 \nonumber \\ 
&& - \int_{\Sigma}{\mathbf E}[\cL_{\zeta_n}\phi, \cL_{\zeta_{-n}}\phi; \bar{\phi}]\Big\}
\Big |_{n^3}
\label{cc}
\end{eqnarray}
where by $|_{n^3}$ we mean that only 
terms proportional to $n^3$ will be extracted, 
and $\mathcal{L}_{\zeta_n}$ denotes the Lie derivative with respect to $\zeta_n$ \eqref{killing}. 
$\mathbf{ X}_{\alpha\beta}$ and $\mathbf{W}^\alpha$ are defined by 
\begin{align}
\begin{split}
\left( \mathbf{ X}_{\alpha\beta} \right)_{\lambda\mu \nu   } 
&= - \epsilon_{\rho \sigma \lambda\mu \nu  }
Z^{\rho\sigma}_{\phantom{\rho \sigma} \alpha\beta}, 
\\
 \left( \mathbf{W}^\alpha \right)_{\lambda\mu \nu  } 
&=
2\left( \nabla_\beta \mathbf{ X}^{\alpha\beta} 
\right)_{\lambda\mu \nu  }.\label{W}
\end{split}
\end{align}

Furthermore, the explicit form of the ${\mathbf E}$-term in \eqref{cc} 
was obtained 
in \cite{0903.4176}; namely, 
\begin{eqnarray}
{\mathbf E}\equiv {\mathbf E}_{\lambda \mu \nu} = \epsilon_{\rho \sigma \lambda \mu \nu}\frac12\left(-\frac{3}{2}Z^{\rho\sigma \gamma \eta} \delta g_{\gamma}{}^{\kappa} \wedge \delta g_{\kappa\eta} + 2Z^{\rho \gamma \eta \kappa} \delta g_{\gamma \eta} \wedge \delta g^{\sigma}{}_\kappa \right),
\label{E}
\end{eqnarray}
where $\phi$ is any field including $g_{\mu \nu}$. Furthermore, ${\mathbf E}[\cL_{\zeta_n}\phi, \cL_{\zeta_{-n}}\phi; \phi]$ is defined as 
\begin{equation}
{\mathbf E}[\cL_{\zeta_n}\phi, \cL_{\zeta_{-n}}\phi; \phi] \equiv (\cL_{\zeta_{-n}}\phi \frac{\partial}{\partial \phi} + \partial_\mu \cL_{\zeta_{-n}}\phi \frac{\partial}{\partial (\partial_\mu \phi)} + \cdots )_{\llcorner} (\cL_{\zeta_{n}}\phi \frac{\partial}{\partial \phi} + \partial_\mu \cL_{\zeta_{n}}\phi \frac{\partial}{\partial(\partial_\mu \phi)} + \cdots )_{\llcorner}{\mathbf E} \label{E-interior} 
\end{equation}
where $\llcorner$ denotes the interior product. From the direct evaluation of \eqref{E-interior} by using \eqref{E}, one explicitly has 
\begin{eqnarray}
{\mathbf E}[\cL_{\zeta_n}\phi, \cL_{\zeta_{-n}}\phi;\phi] &=&\epsilon_{\rho\sigma\lambda \mu \nu}\frac12\Big(-\frac{3}{2}Z^{\rho\sigma \gamma \eta}\{(\cL_{\zeta_{n}}g)_{\gamma}{}^{\kappa}(\cL_{\zeta_{-n}}g)_{\kappa\eta} - (\cL_{\zeta_{n}}g)_{\kappa\eta}(\cL_{\zeta_{-n}}g)_{\gamma}{}^{\kappa}\} \nonumber \\
&&+Z^{\rho \gamma \eta \kappa}\{(\cL_{\zeta_{n}}g)_{\gamma \eta}(\cL_{\zeta_{-n}}g)^{\sigma}{}_{\kappa}-(\cL_{\zeta_{n}}g)^{\sigma}{}_\kappa(\cL_{\zeta_{-n}}g)_{\gamma \eta}\} \Big).\label{EEEE}
\end{eqnarray}
With the expressions \eqref{W} and \eqref{EEEE}, one can explicitly evaluate \eqref{cc}.

Indeed, as shown in \cite{0903.4176}, by 
using the aforementioned definition of $c$, the Iyer-Wald formula 
gets completely reproduced at least in the 4D extremal Kerr cases. 
Here, nevertheless we clarify its validness even for the 5D BMPV black hole. As a matter of fact, it is highly non-trivial to check 
whether \eqref{cc} still holds for the BMPV black hole because its near-horizon topology,  
$S^1$-fibration over 
$AdS_2 \times S^2$, definitely 
differs from that of the 4D extremal Kerr black holes. 

\subsection{Calculation}

Again, let us take the orthonormal basis \eqref{frame2} during  evaluating \eqref{cc} for the sake of computational convenience. 
One can rewrite \eqref{killing} by a basis $e_{\hat{a}} = e_{\hat{a}}{}^{\mu}\partial_{\mu}$ where $e_{\hat{a}}{}^{\mu} = (e^{\hat{a}}{}_{\mu})^{-1}$. Then, the explicit form of $\zeta_{n}$ in terms of the basis $e_{\hat{a}}$ is  
\begin{equation}
\zeta_{n} = -e^{-in\varphi}\left(\frac{\sqrt{\mu-a^2}}{2} e_{\hat{\varphi}}+in\frac{\sqrt{\mu}}{2} e_{\hat{r}}\right).
\label{killing1}
\end{equation}
Equipped with the above \emph{asymptotic} Killing vector \eqref{killing1}, one manages to have 
\begin{eqnarray} 
\Big( \mathcal{L}_{\zeta_n}\nabla^{\hat{a}} \zeta_{-n}^{\hat{b}} \Big) \Big|_{n^3}=
\left(
\begin{array}{ccccc}
0 & \frac{ia}{2\sqrt{\mu-a^2}} & 0 & 0& 0\\
-\frac{ia}{2\sqrt{\mu-a^2}} & 0 & 0& 0 & \frac{i(3a^2\mu - 2\mu^2 + (2a^2\mu-2\mu^2)\cot^2\theta)}{2\mu^{\frac{3}{2}}\sqrt{\mu-a^2}}\\
0 & 0 & 0 & 0& 0\\
0 & 0 & 0 & 0& 0\\
0 & -\frac{i(3a^2\mu - 2\mu^2 + (2a^2\mu-2\mu^2)\cot^2\theta)}{2\mu^{\frac{3}{2}}\sqrt{\mu-a^2}} & 0 & 0& 0
\end{array}
\right)
\label{MMatrix}
\end{eqnarray}
whose columns (rows) are labeled in order by 
$\hat{t}, \hat{r}, \hat{\theta}, \hat{\psi}$ and $\hat{\varphi}$.

By combining both \eqref{MMatrix} and $Z_{\hat{t}\hat{r}\hat{r}\hat{\varphi}}=0$, 
the first term in \eqref{cc} becomes 
\begin{eqnarray}
 c_{{\rm 1st}}&\equiv&-24 i\int_{\Sigma} \left[2Z_{\hat{t}\hat{r}\hat{t}\hat{r}}\left(\frac{ia}{2\sqrt{\mu-a^2}}\right)
-2Z_{\hat{t}\hat{r}\hat{r}\hat{t}}\left(-\frac{ia}{2\sqrt{\mu-a^2}}\right)\right] e^{\hat{\theta}} \wedge e^{\hat{\psi}} \wedge e^{\hat{\varphi}} 
\nonumber\\
&=& 48 \int_{\Sigma}\frac{a}{\sqrt{\mu-a^2}}Z_{\hat{t}\hat{r}\hat{t}\hat{r}}e^{\hat{\theta}} \wedge e^{\hat{\psi}} \wedge e^{\hat{\varphi}}.
\label{c1}
\end{eqnarray} 
Equipped with Cardy's formula \eqref{ca} and the Frolov-Thorne temperature \cite{Azeyanagi:2008dk, 0812.4440} 
\begin{equation} 
T_{FT} = -\frac{\sqrt{\mu - a^2}}{2\pi a},
\label{Frolov-Thorne temperature}
\end{equation} 
we have 
\begin{eqnarray}
S_1 &=& \frac{\pi^2}{3}\int_{\Sigma}\left(48 Z_{\hat{t}\hat{r}\hat{t}\hat{r}} \frac{a}{\sqrt{\mu-a^2}}\right)
\left(-\frac{\sqrt{\mu-a^2}}{2 \pi a}\right)e^{\hat{\theta}} \wedge e^{\hat{\psi}} \wedge e^{\hat{\varphi}} \nonumber \\
&=&-8\pi\int_{\Sigma}Z_{\hat{t}\hat{r}\hat{t}\hat{r}}e^{\hat{\theta}} \wedge e^{\hat{\psi}} \wedge e^{\hat{\varphi}}.
\label{eq:central_charge}
\end{eqnarray}
It precisely coincides with the Iyer-Wald result \eqref{wald1}. Surely, we have assumed 
the validness of the Frolov-Thorne temperature even in the presence of  the $R^2$-curvature corrections.%
\footnote{
It was argued in \cite{0903.4176} that 
higher-derivative corrections have no effect on 
the Frolov-Thorne temperature of 4D 
extremal Kerr black holes.} 
In view of 
\eqref{Frolov-Thorne temperature}, it seems that when $a>0$ the above 
$T_{FT}$ of some $putative$ 2D CFT become negative. This is a quite common issue encountered during applying Kerr/CFT correspondence. 
The sign dependence on $a$ is merely an illusion because it has no real effect on a physical quantity like entropy $S_1$ which is independent of the sign of $a$. 
Certainly, another choice of $\zeta$ associated with $\psi$ is possible but leads to a vanishing central charge. 
This situation which resembles the tree-level case 
encountered in \cite{0812.4440} may be attributable to 
zero angular momentum along $\psi$. 

Let us examine other contributions to the central charge $c$ in \eqref{cc}. We will see that 
they cancel one another out eventually. 
Since we have already reproduced the Iyer-Wald result only from the first term of \eqref{cc}, the contributions from the other terms have to cancel one another out. 
In \eqref{cc}, the explicit evaluation can show that the second and third terms as a whole give
\begin{eqnarray}
-24 i  \int_{\Sigma}  
\left[\left( \mathcal{L}_{\zeta_n} \mathbf{ X}\right)_{\alpha\beta}
\nabla^{[\alpha} \zeta^{\beta]}_{-n}
+ \mathcal{L}_{\zeta_n} \mathbf{W}_\alpha \zeta^{\alpha}_{-n}
\right]\Big|_{n^3} 
= 384\frac{a\sqrt{\mu-a^2}}{\mu^2}\int_{\Sigma}e^{\hat{\theta}} \wedge e^{\hat{\psi}} \wedge e^{\hat{\varphi}}.
\label{cancel1}
\end{eqnarray}
On the other hand, one can explicitly compute 
\eqref{EEEE} via the near-horizon metric \eqref{eq:BMPV_horizon}, and the result is
\begin{equation}
-12 i \int_{\Sigma}{\mathbf E}[\cL_{\zeta_n}\phi, \cL_{\zeta_{-n}}\phi;\bar{\phi}] \Big|_{n^3}=-384\frac{a\sqrt{\mu-a^2}}{\mu^2}\int_{\Sigma}e^{\hat{\theta}} \wedge e^{\hat{\psi}} \wedge e^{\hat{\varphi}}.
\label{cancel2}
\end{equation}
Consequently, a perfect cancellation happens as expected. Namely, we have confirmed that the entropy obtained from the central charge \eqref{cc} precisely reproduces the Iyer-Wald formula \eqref{S1}.

\section{Summary}
\label{sec:sum}

Our results are summarized as follows. 
Plugging into Cardy's formula of 2D CFT \eqref{ca} the central charge spelt out in 
\cite{0111246, 0708.2378, 0708.3153} and Frolov-Thorne temperature analyzed in 
\cite{Azeyanagi:2008dk, 0812.4440}, we obtained the entropy of BMPV black holes when 
Gauss-Bonnet terms are present. 
This computation, though semiclassical, can be regarded as a microscopic derivation in contrast to Iyer-Wald formula. 

Because there are two independent $(\varphi, \psi)$, it is $\varphi$-direction associated 
with non-zero angular momentum that gives us the finite 
central charge. This resembles much the tree-level 
(Einstein-Hilbert action) situation 
encountered in \cite{0812.4440}. It will be interesting to understand this 
phenomenon further within a more general framework.


\section*{Acknowledgments}
We thank Chiang-Mei Chen, Kevin Goldstein, Hiroshi Isono, Yoshinori Matsuo, 
Hiroaki Nakajima, Noriaki Ogawa, Hesam Soltanpanahi and Wen-Yu Wen for discussions. 
In particular, we are grateful to Yuji Tachikawa for explaining 
us arXiv:0903.4176. 
TST owes Yutaka Baba for his early collaboration. 
He also thanks Takashi Okamura and Masato Minamitsuji for their hospitality.  

\appendix

\end{document}